\documentclass[%
	conference,
]{IEEEtran}

\usepackage{graphicx}
\usepackage{epstopdf}
\usepackage{standalone}
\usepackage[nolist ]{acronym}
\usepackage{tcolorbox}

\usepackage{pdfpages}
\usepackage{tikz}
\usetikzlibrary{positioning}
\usepackage[bookmarks=false]{hyperref}
\usepackage{pdfcomment}
\usepackage{hologo}

\usepackage{xstring}

\usepackage[utf8]{inputenc}
\usepackage{marvosym}

\usepackage{algorithm}
\usepackage{algpseudocode}
\usepackage{tikz}

\usepackage{listings}
\definecolor{ListingBackground}{rgb}{0.97,0.97,0.97}
\usepackage{pgfplots}
\pgfplotsset{compat=newest}
\pgfplotsset{
	box plot/.style={
		/pgfplots/.cd,
		fill=blue!30,
		only marks,
		mark=-,
		mark size=0.2em,
		/pgfplots/error bars/.cd,
		y dir=plus,
		y explicit,
	},
	box plot box/.style={
		/pgfplots/error bars/draw error bar/.code 2 args={%
			\draw  ##1 -- ++(.2em,0pt) |- ##2 -- ++(-.2em,0pt) |- ##1 -- cycle;
		},
		/pgfplots/table/.cd,
		y index=2,
		y error expr={\thisrowno{3}-\thisrowno{2}},
		/pgfplots/box plot
	},
	box plot top whisker/.style={
		/pgfplots/error bars/draw error bar/.code 2 args={%
			\pgfkeysgetvalue{/pgfplots/error bars/error mark}%
			{\pgfplotserrorbarsmark}%
			\pgfkeysgetvalue{/pgfplots/error bars/error mark options}%
			{\pgfplotserrorbarsmarkopts}%
			\path ##1 -- ##2;
		},
		/pgfplots/table/.cd,
		y index=4,
		y error expr={\thisrowno{2}-\thisrowno{4}},
		/pgfplots/box plot
	},
	box plot bottom whisker/.style={
		/pgfplots/error bars/draw error bar/.code 2 args={%
			\pgfkeysgetvalue{/pgfplots/error bars/error mark}%
			{\pgfplotserrorbarsmark}%
			\pgfkeysgetvalue{/pgfplots/error bars/error mark options}%
			{\pgfplotserrorbarsmarkopts}%
			\path ##1 -- ##2;
		},
		/pgfplots/table/.cd,
		y index=5,
		y error expr={\thisrowno{3}-\thisrowno{5}},
		/pgfplots/box plot
	},
	box plot median/.style={
		/pgfplots/box plot
	},
	boxplot/every median/.style={
		ultra thick,dashed,cyan
	}
}

\definecolor{flexicolor}{RGB}{46,49,146}
\definecolor{amaricolor}{RGB}{237,28,36}

\usepackage{xspace}

\usepackage[binary-units=true]{siunitx}
\usepackage{psfrag}
\usepackage{graphicx}
\usepackage{tabularx,booktabs}
\usepackage{multirow}
\usepackage{rotating}

\renewcommand{\baselinestretch}{1}

\usepackage{multirow}

\usepackage[cmex10]{amsmath}
\usepackage[caption=false,font=footnotesize]{subfig}
\usepackage{stfloats}
\begin{document}

\newcommand{\paperTitle}{System-of-Systems Modeling, Analysis and Optimization of Hybrid Vehicular Traffic}
\newcommand{\paperAuthors}{Benjamin Sliwa and Christian Wietfeld}
\newcommand{\paperEmails}{$\{$Benjamin.Sliwa, Christian.Wietfeld$\}$@tu-dortmund.de}

\newcommand{\figurePadding}{0pt}
\newcommand{\figureTopPadding}{\figurePadding}
\newcommand{\figureBottomPadding}{\figurePadding}

\newcommand{\dummy}[3]
{
	\begin{figure}[b!]  
		\begin{tikzpicture}
		\node[draw,minimum height=6cm,minimum width=\columnwidth]{\LARGE #1};
		\end{tikzpicture}
		\caption{#2}
		\label{#3}
	\end{figure}
}

\newcommand{\wDummy}[3]
{
	\begin{figure*}[b!]  
		\begin{tikzpicture}
		\node[draw,minimum height=6cm,minimum width=\textwidth]{\LARGE #1};
		\end{tikzpicture}
		\caption{#2}
		\label{#3}
	\end{figure*}
}

\newcommand{\basicFig}[7]
{
	\begin{figure}[#1]  	
		\vspace{#6}
		\centering		  
		\includegraphics[width=#7\columnwidth]{#2}
		\caption{#3}
		\label{#4}
		\vspace{#5}	
	\end{figure}
}
\newcommand{\fig}[4]{\basicFig{#1}{#2}{#3}{#4}{0cm}{0cm}{1}}

\newcommand{\subfig}[3]
{
	\subfloat[#3]
	{
		\includegraphics[width=#2\textwidth]{#1}
	}
	\hfill
}

\newcommand\circled[1] 
{
	\tikz[baseline=(char.base)]
	{
		\node[shape=circle,draw,inner sep=1pt] (char) {#1};
	}\xspace
}
\begin{acronym}
	\acro{LIMoSim}{Lightweight ICT-centric Mobility Simulation}
	\acro{CAT}{Channel-aware Transmission}
	\acro{pCAT}{predictive CAT}
	\acro{ML-CAT}{Machine Learning-based CAT}
	\acro{ML-pCAT}{Machine Learning-based pCAT}
	\acro{MTC}{Machine-type Communication}
	
	\acro{LTE}{Long Term Evolution}
	\acro{UE}{User Equipment}
	\acro{RSRP}{Reference Signal Received Power}
	\acro{RSRQ}{Reference Signal Received Quality}
	\acro{SINR}{Signal-to-noise-plus-interference Ratio}
	\acro{CQI}{Channel Quality Indicator}
	\acro{V2X}{Vehicle-to-everything}
	
	\acro{ITS}{Intelligent Transportation System}
	\acro{QUIC}{Quick UDP Internet Connections}
	\acro{mmWave}{millimeter Wave}
	\acro{RSU}{Road Side Unit}
	\acro{SVM}{Support Vector Machine}
	\acro{xFCD}{extended Floating Car Data}
	\acro{RSSI}{Received Signal Strength Indicator}
	\acro{IDM}{Intelligent Driver Model}
	\acro{BMP}{Breakdown Minimization Principle}
	\acro{WE}{Wardrops User Equilibrium}
\end{acronym}

\newcommand\lte{\ac{LTE}\xspace}
\newcommand\ue{\ac{UE}\xspace}
\newcommand\limosim{\ac{LIMoSim}\xspace}
\newcommand\cat{\ac{CAT}\xspace}
\newcommand\pcat{\ac{pCAT}\xspace}
\newcommand\mlcat{\ac{ML-CAT}\xspace}
\newcommand\mlpcat{\ac{ML-pCAT}\xspace}
\newcommand\mtc{\ac{MTC}\xspace}
\newcommand\rsrp{\ac{RSRP}\xspace}
\newcommand\rsrq{\ac{RSRQ}\xspace}
\newcommand\sinr{\ac{SINR}\xspace}
\newcommand\cqi{\ac{CQI}\xspace}
\newcommand\its{\ac{ITS}\xspace}
\newcommand\quic{\ac{QUIC}\xspace}
\newcommand\mmWave{\ac{mmWave}\xspace}
\newcommand\rsu{\ac{RSU}\xspace}
\newcommand\rsus{\acp{RSU}\xspace}
\newcommand\svm{\ac{SVM}\xspace}
\newcommand\xfcd{\ac{xFCD}\xspace}
\newcommand\rssi{\ac{RSSI}\xspace}
\newcommand\idm{\ac{IDM}\xspace}
\newcommand\bmp{\ac{BMP}\xspace}
\newcommand\we{\ac{WE}\xspace}
\acresetall
\title{\paperTitle}

\newcommand\Mark[1]{\textsuperscript#1}

\author{
	
	\IEEEauthorblockN{
		\textbf{Benjamin Sliwa\Mark{1}, Thomas Liebig\Mark{2}, Tim Vranken\Mark{3}, Michael Schreckenberg\Mark{3} and Christian Wietfeld\Mark{1}}
	}

	\IEEEauthorblockA{
		\Mark{1}Communication Networks Institute, 
		\Mark{2}Department of Computer Science VIII\\ 
		TU Dortmund University, 44227 Dortmund, Germany\\
		e-mail:  $\{$Benjamin.Sliwa, Thomas.Liebig, Christian.Wietfeld$\}$@tu-dortmund.de\\
		\Mark{3}Physics of Transport and Traffic, University Duisburg-Essen, Germany\\
		e-mail:  $\{$Vranken, Schreckenberg$\}$@ptt.uni-due.de 	
	}
}

\maketitle

%
%
\def\COPYRIGHTYEAR{2019}
\def\CONFERENCE{2019 Annual IEEE International Systems Conference (SysCon)} 
\def\DOI{10.1109/SYSCON.2019.8836786}	
\def\bibtex
{
	@InProceedings\{Sliwa/etal/2019b,
	author    = \{Benjamin Sliwa and Thomas Liebig and Tim Vranken and Michael Schreckenberg and Christian Wietfeld\},
	title     = \{System-of-systems modeling, analysis and optimization of hybrid vehicular traffic\},
	booktitle = \{2019 Annual IEEE International Systems Conference (SysCon)\},
	year      = \{2019\},
	address   = \{Orlando, Florida, USA\},
	month     = \{Apr\},
	publisher = \{IEEE\},
	\}
}
\ifx\CONFERENCE\VOID
\def\conferencenotice{Submitted for publication}
\def\copyrightnotice{}
\else
\ifx\DOI\VOID
\def\conferencenotice{Accepted for presentation in: \CONFERENCE}	
\else
\def\conferencenotice{Published in: \CONFERENCE\\DOI: \href{http://dx.doi.org/\DOI}{\DOI}

}
\fi
\def\copyrightnotice{
	\copyright~\COPYRIGHTYEAR~IEEE. Personal use of this material is permitted. Permission from IEEE must be obtained for all other uses, including reprinting/republishing this material for advertising or promotional purposes, collecting new collected works for resale or redistribution to servers or lists, or reuse of any copyrighted component of this work in other works.
}
\fi
\def\overlayimage{%
	\begin{tikzpicture}[remember picture, overlay]
	\node[below=5mm of current page.north, text width=20cm,font=\sffamily\footnotesize,align=center] {\conferencenotice \vspace{0.3cm} \pdfcomment[color=yellow,icon=Note]{\bibtex}};
	\node[above=5mm of current page.south, text width=15cm,font=\sffamily\footnotesize] {\copyrightnotice};
	\end{tikzpicture}%
}
\overlayimage
\begin{abstract}
	
%
%
While the development of fully autonomous vehicles is one of the major research fields in the \acp{ITS} domain, the upcoming long-term transition period -- the hybrid vehicular traffic -- is often neglected.
%
%
However, within the next decades, automotive systems with heterogeneous autonomy levels will share the same road network, resulting in new problems for traffic management systems and communication network infrastructure providers.
%
%
In this	paper, we identify key challenges of the upcoming hybrid traffic scenario and present a system-of-systems model, which brings together approaches and methods from traffic modeling, data science, and communication engineering in order to allow data-driven traffic flow optimization.
%
%
The proposed model consists of data acquisition, data transfer, data analysis, and data exploitation and exploits real world sensor data as well as simulative optimization methods. Based on the results of multiple case studies, which focus on individual challenges (e.g., resource-efficient data transfer and dynamic routing of vehicles), we point out approaches for using the existing infrastructure with a higher grade of efficiency.

\end{abstract}

\IEEEpeerreviewmaketitle

\section{Introduction and Related Work} \label{sec:introduction}

%
%
Autonomous vehicles are often envisioned as the savior that solves all issues \cite{Brenner/Herrmann/2018a} of current traffic systems. They are expected to be able to drive with minimum safety distances even at high speeds, which enables massive increases of the usable road capacity. They do not dawdle, which results in a better traffic flow and they coordinate their actions for an improved overall safety.

%
%
While this sophisticated vision is supported by many current research works \cite{Calafate/etal/2017a}, most of them neglect the more immediate challenges arising from the transition period from legacy traffic to highly automated vehicular traffic that will take several decades. Due to the step-wise development and deployment of automated vehicles, systems with different levels of autonomy will coexist within the same road network in the \emph{hybrid vehicular traffic} \cite{International/2014a}.
%
%
In this complex scenario, the following challenges for traffic flow optimization arise:
%
%
\begin{itemize}
	\item \textbf{Novel traffic phenomenons} as automated vehicles are expected to react defensively on human misbehavior, potentially resulting in traffic flow reductions, while the overall traffic congestion grows, e.g., due to an increased amount of empty runnings and on-demand traffic.
	\item \textbf{Heterogeneous communication technologies} limit the capabilities for mutual coordination and data provisioning.
	\item \textbf{Resource competition} as the radio spectrum is shared between the different technologies and human cell users as well as automated systems.
	\item Traffic management systems have to work with \textbf{incomplete information}, since established sensors systems only provide local coverage and often break during their operational time.
\end{itemize}
%
%
\fig{b}{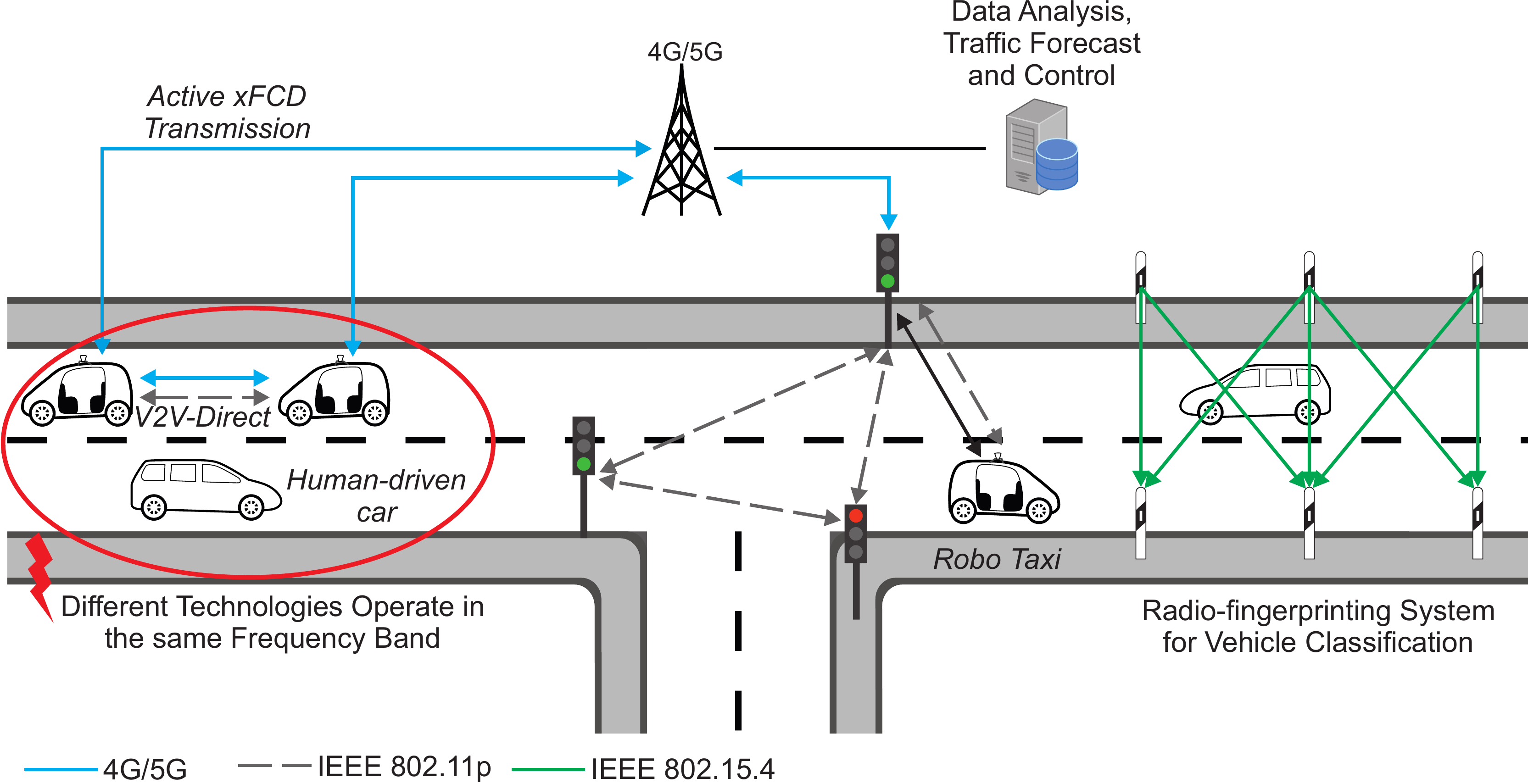}{Hybrid vehicular traffic: Coexistence of different vehicle types and communication technologies.}{fig:scenario}
Existing approaches for modeling these traffic systems have not yet reached the required complexity that allows the derivation of novel insights for hybrid traffic flow optimization. In \cite{Ye/Yamamoto/2018a}, the authors propose an approach to model different automation levels depending on their resulting safety distance. Another approach \cite{Zhou/etal/2017a} proposes an extension of the well-established \idm to consider cooperative behaviors for lane-merging.
While forecast and optimization of vehicular traffic flow has traditionally been isolatedly addressed by the traffic physics community, the arising challenges of the hybrid traffic scenario require the joint consideration of communication systems and data analysis methods \cite{Chen/etal/2017a}.
%
%
In this paper, we present an overall system-of-systems model that brings together the three mentioned disciplines for modeling and analyzing hybrid vehicular traffic with the goal of data-driven traffic flow optimization. The concepts are substantiated by preliminary results of different case studies that focus on individual aspects such as resource-efficient data transfer and traffic bottleneck detection. The proposed approach leverages ubiquitous machine learning to derive novel insights within all of its main dimensions, which are defined as data acquisition, data transfer, data analysis and data exploitation.

Fig.~\ref{fig:scenario} shows an example scenario and illustrates the different kinds of heterogeneity within hybrid traffic scenarios. While highly automated vehicles (so-called \emph{robo taxis}) are exploited as moving sensor nodes that provide sensor measurements as \xfcd via different kinds of \ac{V2X} connections on their own, legacy vehicles are detected and classified based on roadside infrastructure systems.

%
%
The structure of the paper is as follows. After presenting the challenges and discussing relevant state-of-the-art research, we present the architecture model of the proposed overall system. The different submodules are explained in details and set into relation to the overall context and existing approaches. Finally, multiple case-studies that provide groundwork for the different components of the overall system are presented and selected results are discussed.

\section{System Model Architecture} \label{sec:approach}

In this section, the system architecture and the features of the individual components are explained.
%
%
\begin{figure*}
	\includegraphics[width=1\textwidth]{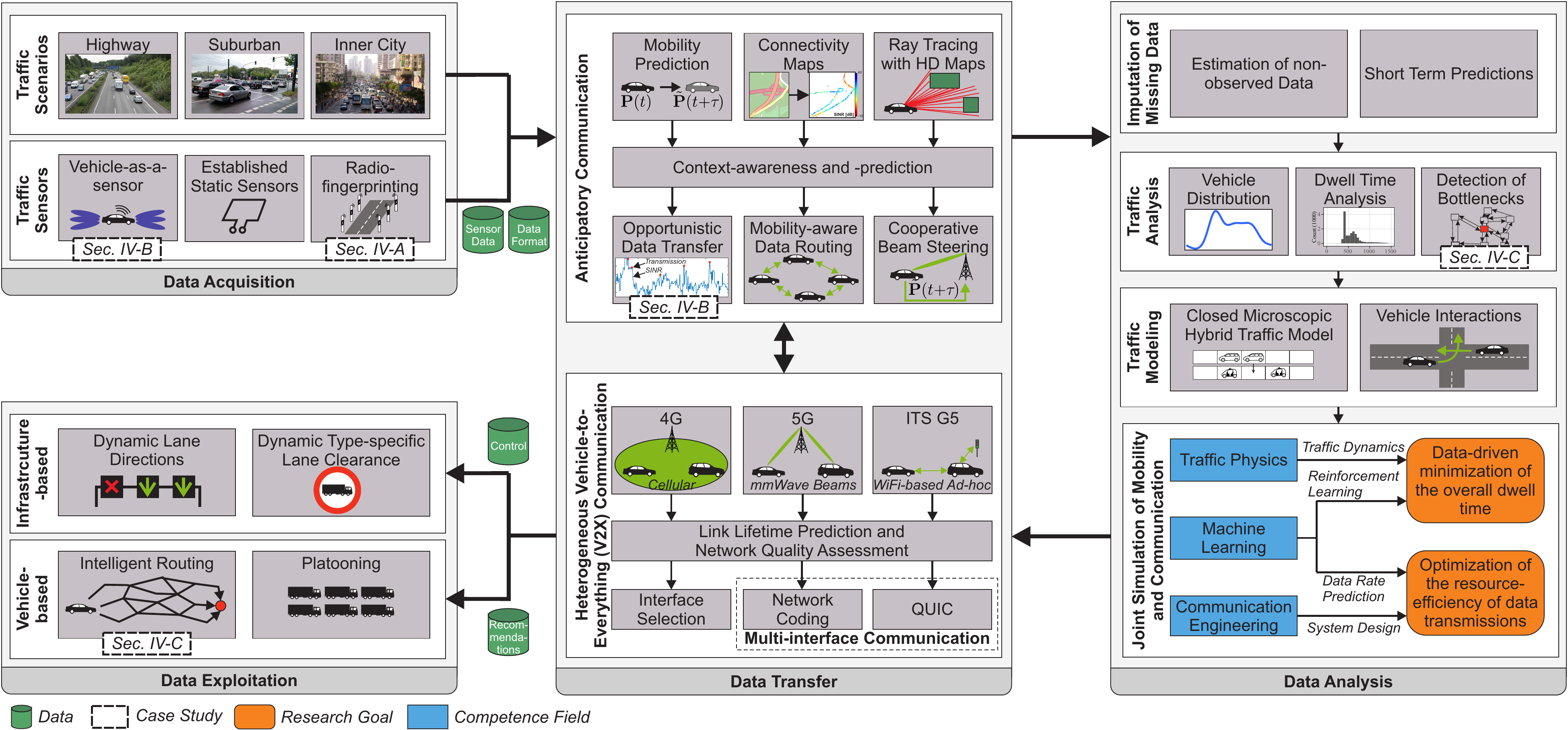}
	\label{fig:system_architecture}
	\caption{Overall system architecture model for data-driven optimization of hybrid traffic consisting of data acquisition, data transfer, data analysis and data optimization. Real world sensor data is exploited to generate new traffic models, which are evaluated and optimized using a simulation setup. The derived insights are then utilized for traffic flow optimization in real world traffic management systems.}
\end{figure*}
An illustration of the overall system architecture model is shown in Fig.~\ref{fig:system_architecture}. For the proposed data-driven approach, traffic data is acquired in the real world, transfered to the analysis centers, utilized for model generation and simulative optimization and finally exploited for improving the real world traffic flow.
In the overall scenario, communication systems and machine learning-based data analysis are of great importance. One the one hand, communication is used as a mediator between data acquisition, data analysis and data exploitation, on the other hand, it is subject to machine learning-based optimization itself (cf. Sec.~\ref{sec:cat}).
The following paragraphs provide further information about exemplary groundwork for the key system stages \emph{data acquisition}, \emph{data transfer}, \emph{data analysis} and \emph{data exploitation}. In addition, groundwork that addresses the related challenges is introduced and discussed.

\subsection{Data Acquisition}

The derivation of new models and the following traffic flow optimization requires real world traffic data as a solid foundation. Therefore, the accurate acquisition of traffic data is of tremendous importance for the overall system.
%
%
Depending on the considered road network type, the degrees of freedom for the vehicular mobility and the resulting challenges for the traffic flow analysis can be significantly different. Consequently, traffic is analyzed isolatedly for highway, suburban and inner city scenarios as well as in an overall model, which considers all traffic types.
%
%
For the acquisition of actual sensor data, a combination of different real world sensor systems is applied, which includes existing traditional sensors as well as experimental approaches that offer novel acquisition methods and are capable of opening up new data categories.
Traditional sensor systems (e.g., induction loops) are intended to acquire traffic parameters such as traffic flow and density. Due to the required road work, sensor deployments are cost-intense, which results in limitations for the number of deployments and the resulting sensor coverage. However, since many existing traffic monitoring systems rely on these approaches, they are integrated into the system model in order to benefit from the existing infrastructure.
%
%
In addition to these established sensor systems, novel methods allow the acquisition of additional traffic parameters (e.g., vehicle type distributions), which enables new approaches for traffic flow optimization.

\subsubsection{Vehicle-as-a-sensor}

Vehicular crowdsensing \cite{Wang/etal/2016a} marks a paradigm shift from static sensor installations to exploiting the vehicles themselves as moving sensor nodes, resulting in massive improvements for coverage and data freshness. Using a multitude of specialized sensors, vehicles can be utilized to provide meaningful traffic parameters as well as environmental information, e.g., to detect traffic obstructions or free parking slots.
%
%
Apart from the considered traffic management scenario, this approach has been demonstrated to significantly improve distributed weather prognosis \cite{Calafate/etal/2017a}, road condition sensing \cite{Wang/etal/2017a} and inner city air quality monitoring.

%
%
Vehicular crowdsensing is closely connected to the \emph{data transfer} topic, as it relies on connectivity and the presence of a data uplink. Moreover, large-scale deployments of crowdsensing services lead to massive increases in the amount of transmitted data, which imply novel challenges on the network infrastructure side (see Sec.~\ref{sec:cat}).

\subsubsection{Novel Roadside Sensors}

In order to enable promising traffic optimization methods like type-specific lane clearance, traffic management systems need to obtain information about the distributions of vehicle types.
A multitude of different sensor types for vehicle classification has been proposed, which ranges from vision-based approaches using cameras, radars and lasers to inertial measuring units. A comprehensive overview about existing sensor technologies is presented in \cite{Guerrero-Ibanez/etal/2018a}. However, none of the existing technologies is able to fulfill all requirements for accuracy, privacy preservation, weather robustness while still being cost-efficiently enough to allow citywide large-scale deployments.

%
%
In Sec.~\ref{sec:ffws}, we present a case study on radio-fingerprinting-based vehicles classification. The proposed system can be deployed cost-efficiently in an ad-hoc manner and is able to provide a high classification accuracy even at challenging environmental conditions.

\subsection{Data Transfer} \label{sec:data_transfer}

Alongside with the coexistence of the different autonomy levels comes the coexistence of different communication technologies, that compete over the available resources in the shared radio medium. Within the proposed system for hybrid vehicular traffic analysis, data needs to be exchanged from the cars to the cloud-based data analytics center and between individual vehicles for mutual coordination.
The challenges arising from highly-dynamic network topologies, heterogeneous communication technologies and challenging environments are addressed by the combined usage of two main methodological approaches: \emph{Anticipatory communication} and \emph{heterogeneous \ac{V2X}}.

\subsubsection{Anticipatory Communication} \label{sec:car_to_cloud}

The anticipatory networking paradigm \cite{Bui/etal/2017a} aims to integrate the communication context into the decision processes within communication systems themselves. Instead of extending the network capacity physically on the infrastructure side, the existing network itself is used in a more efficient way through a context-aware medium access.
%
%
In the considered vehicular scenario, the dynamics of the communication medium are closely connected to the vehicles mobility behavior. Therefore, for forecasting the network quality a car is going to encounter in the future, trajectory prediction is a basic requirement \cite{Sliwa/etal/2018a}.

%
%
For improving the coexistence of different resource-consuming entities, the resource-efficiency of their individual data transmissions needs to be optimized. Resource-efficient communication has several dimensions: Data transmissions need to be short (high data rates) in order to free the occupied resources as soon as possible, they have to be reliable (low amount of retransmissions) and they have to satisfy the data freshness requirements of the intended application.
%
%
Although different indicators exist for assessing the current network quality of a given communication technology, the validity of the indicators is related to the channel coherence time and the transmission patterns of the application itself. Machine learning has been demonstrated \cite{Sliwa/etal/2016b} to be able to efficiently consider hidden interdependencies between those aspects, resulting in an improved context-awareness. In Sec.~\ref{sec:cat}, we present a case study for resource-efficient sensor data transmission based on machine learning and mobility prediction.

%
%
In addition to optimizing the resource-efficiency of data transmissions, anticipatory communication can serve as a method to establish robust communication paths within vehicular mesh networks. Here, the consideration of the mobility characteristics of the vehicles allows to derive link-lifetime estimations that enables the predictive optimization of multi-hop routes \cite{Sliwa/etal/2016b}.
%
%
With the deployments of 5G communication networks and the \mmWave technology, context-aware networking becomes a native aspect of mobile networking, since the extremely high directed radio beams need to be precisely steered towards their moving receiver.

\subsubsection{Heterogeneous \ac{V2X}}

As it is expected that vehicles will make use of a combination of different technologies for specific use-cases \cite{Cavalcanti/etal/2018a} -- e.g., low-latency ad-hoc communication for safety-related messaging and high data rate cellular communication for sensor data transfer -- the unique technological properties can be exploited for dynamic technology selection. Moreover, for each data transfer, the currently used communication technology is chosen in a context-aware manner, e.g., available \rsus are opportunistically exploited as cost-efficient internet gateways.
Anticipatory communication methods are applied to chose the respective technology in a predictive manner with respect to the application requirements, e.g., through prediction of link availability times.
In addition, multi-interface communication using network coding \cite{Liu/etal/2016a} and \quic is applied in order to utilize multiple links in parallel.

\subsection{Data Analysis}

In the context of the proposed overall system, data analysis is utilized for simulative traffic flow optimization as well as for the generation of hybrid traffic models themselves. Before the required traffic properties are analyzed, the acquired sensor data is utilized to derive estimations about the traffic situation at non-observed measurement locations in order to obtain an improved awareness of the overall system dynamics.

\subsubsection{Imputation of Missing Data}
%
%
Traffic volume estimation is a fundamental task in macroscopic street-based traffic analysis systems and
has important applications, e.g., quality-of-service evaluation, location evaluation or risk
analysis. Nowadays, intelligent transportation systems rely on stationary sensors, which provide
traffic volume measurements at predefined locations. However, imputation of the unobserved traffic
flow values and short-term predictions are highly important research topics.

Existing literature distinguishes between average daily traffic (ADT) estimation
and average annual daily traffic flow (AADT or AADF) estimation. Whereas AADF focuses on
estimation of a traffic volume depending on the day of the year, ADT estimation
provides an average for a particular day.
In contrast to the ADT and AADF problem, also short-term traffic prognosis
problems exist. In AADT, we concentrate on more extended temporal resolutions, where microscopic influences, e.g., signals
have no impact on the traffic flow. For short-term prediction problems, we point the interested
reader on related works, e.g., Cellular Automaton \cite{nagel/1992}, Poisson Dependency Networks \cite{Habel/etal/2016a}, or Convolutional Neural Networks \cite{ma/2017}.

A naive approach for AADF estimation is the utilization of ordinary linear regression (OLR) \cite{zhao/2004}. Street segment attributes (e.g.
the number of lanes or function classes) are multiplied by weights, which are subject
for least squares regression. Improvements of this technique were achieved by
respecting the geographical space by usage of geographically weighted regression
(GWR) \cite{zhao/2004} and by application of k-nearest neighbor approaches (kNN) \cite{gong/2002}.
In \cite{lam/2006} the AADF prediction of kNN for a particular location is improved by weighting
measurements by their temporal distance to the prediction time. This approach
showed better results than the application of Gaussian maximum likelihood (GML)
approaches for weighting of the historical data points.

Being a spatial regression problem, usage of the Kriging method seems to be a natural choice to tackle
the AADF problem. This was successfully carried on at University of Texas
\cite{selby/2011,selby/2011a}. To machine learning persons the Kriging method is better known as
Gaussian Process Regression, which allows a better understanding of the underlying spatial
correlation model by reformulation with a kernel matrix. Application of
Gaussian Process Regression is an appealing state-of-the-art method that outperforms recent
methods \cite{liebig/2013}. The method bases on a covariance matrix that denotes the
correlations among the traffic flux values at various locations. The work in \cite{liebig/2013}
tested various covariance matrices among them some that incorporate spatial layout of the sensor
locations or even the topology of the street network. However, the performance did not change much
for these different correlation models. However, due to the computational complexity
of Gaussian Process Regression, application to urban areas was restricted either to small sites or
a sample of locations \cite{artikis/2014}.

\subsubsection{Traffic Modeling and Analysis}
 
Although a wide range of different macro- and microscopical traffic models exist, a closed model for hybrid vehicular traffic is still missing. Therefore, the insights of the data acquisition and the data transfer stages shall be brought together to extend existing cellular automaton models \cite{Knospe/etal/2000a} to consider different levels of automation.

\subsubsection{Integrated Joint Simulation of Mobility and Communication}

As pointed out in the previous paragraphs, modeling hybrid vehicular traffic requires the consideration of traffic modeling, data analysis and communication. For the simulative evaluation and optimization, a common simulation setup that is able to consider all of the individual aspects needs to be created.
%
%
Current approaches for simulating vehicular networks often rely on coupling specialized simulation frameworks based on an interprocess-communication approach \cite{Sommer/etal/2011a}. As a result, the access to information of a specific domain (e.g., using mobility information from the traffic simulator for decisions in the network simulator) is complicated, as specified interfaces and communication protocols are used for the data exchange.
%
%
In contrast to that, the proposed system model relies on an integrated approach based on \limosim \cite{Sliwa/etal/2017b}, which makes use of a shared codebase approach for all different modules.

\subsection{Data Exploitation}

Although a naive approach to increase the road capacity is to construct additional road infrastructure, it is highly cost-intense and often not even applicable due to space limitation, especially in inner city scenarios. Therefore, a more preferable way is to utilize the existing road network in a more efficient way. In the proposed overall system, this approach is addressed by \emph{infrastructure-} and \emph{vehicle based methods}, which exploit the results of the previous data analysis stage.

\subsubsection{Infrastructure-based Methods}

Infrastructure-based methods are centralized approaches, therefore the traffic management system is able to send control commands directly to the corresponding road infrastructure, which then executes the intended action.
%
%
Assigning the directionality of lanes with respect to the capacity requirements provides a method for better coping with temporal-limited phenomenons (e.g., commuter traffic) of traffic systems.
%
%
Another optimization approach is to use lanes exclusively for certain vehicle types. Based on the obtained knowledge about the distribution of vehicle types, lanes can be dynamically assigned, e.g., for dividing passenger cars and heavy vehicles. Moreover, the same approach can be applied for dividing non-automated and highly-automated vehicles, which reduces the need for cross-automation-level coordination and enables traffic systems to benefit more from the advantages of automated vehicles.

\subsubsection{Vehicle-based Methods}

Vehicle-based methods are decentralized approaches, therefore traffic management systems are only able to send recommendations to the vehicles, which then may react accordingly or not. Moreover, in the considered hybrid traffic scenario not all vehicles are equipped with communication technology.

%
%
Autonomous cars can also leverage the vast amount of data in our scenario resulting from sensors, imputations and predictions and apply these data for short-term navigation decisions, i.e., which direction to chose at a particular junction for not to generate or participate in a traffic congestion. This routing optimization approach was tested with bandit learning in \cite{Liebig/etal/2017a} and achieved increased traffic network performance. In \cite{Vranken/etal/2018a}, multiple trip planing methods, which aim to better distribute the traffic flow over the system are investigated. The case study in Sec.~\ref{sec:simu} provides a methodological summary of the mentioned paper.

%
%
Another promising method for traffic flow optimization, which is especially interesting for freight traffic, is the usage of platooning, where vehicles coordinate their velocity characteristics to achieve minimal safety distances. Additional benefits are expected for combining platooning methods with exclusive lanes for platoons.

\section{Groundwork and Case Studies}

In this section, we present multiple case studies that provide groundwork for isolated research fields in the considered overall system model.

\subsection{Data Acquisition: Radio-fingerprinting-based Vehicle Classification} \label{sec:ffws}

In this case study, we present example results of our research work on radio-fingerprinting-based vehicle classification. The integration of class-specific traffic analysis allows to utilize promising traffic flow optimization techniques such as dynamic type-specific lane clearance. The proposed system aims to provide a highly-accurate low-cost alternative to existing sensor systems.
Within the considered hybrid traffic scenario, the system closes the gap between connected vehicles, which are able to provide class information on their own, and non-connected vehicles, which have to be detected using roadside infrastructure systems.
In \cite{Sliwa/etal/2018e}, we have proposed initial work for a radio-fingerprinting system, which leverages the idea of \emph{radio tomography} for vehicle classification and can be cost-efficiently deployed in an ad-hoc manner without requiring additional roadwork.
The proposed system consists of low-cost communicating sensor nodes that are installed into delineator posts. Three transmitters continuously send data to three receivers on the other road side, which monitor the \rssi of all links. Vehicles passing the sensor installation attenuate the signals, whereas the attenuation level and the duration of the attenuation phase is related to the vehicle shape. Therefore, the time series values of all nine links are utilized as a vehicle class-specific \emph{radio fingerprint}. The attenuation pattern itself is utilized for machine learning-based vehicle classification.
%
%
\fig{}{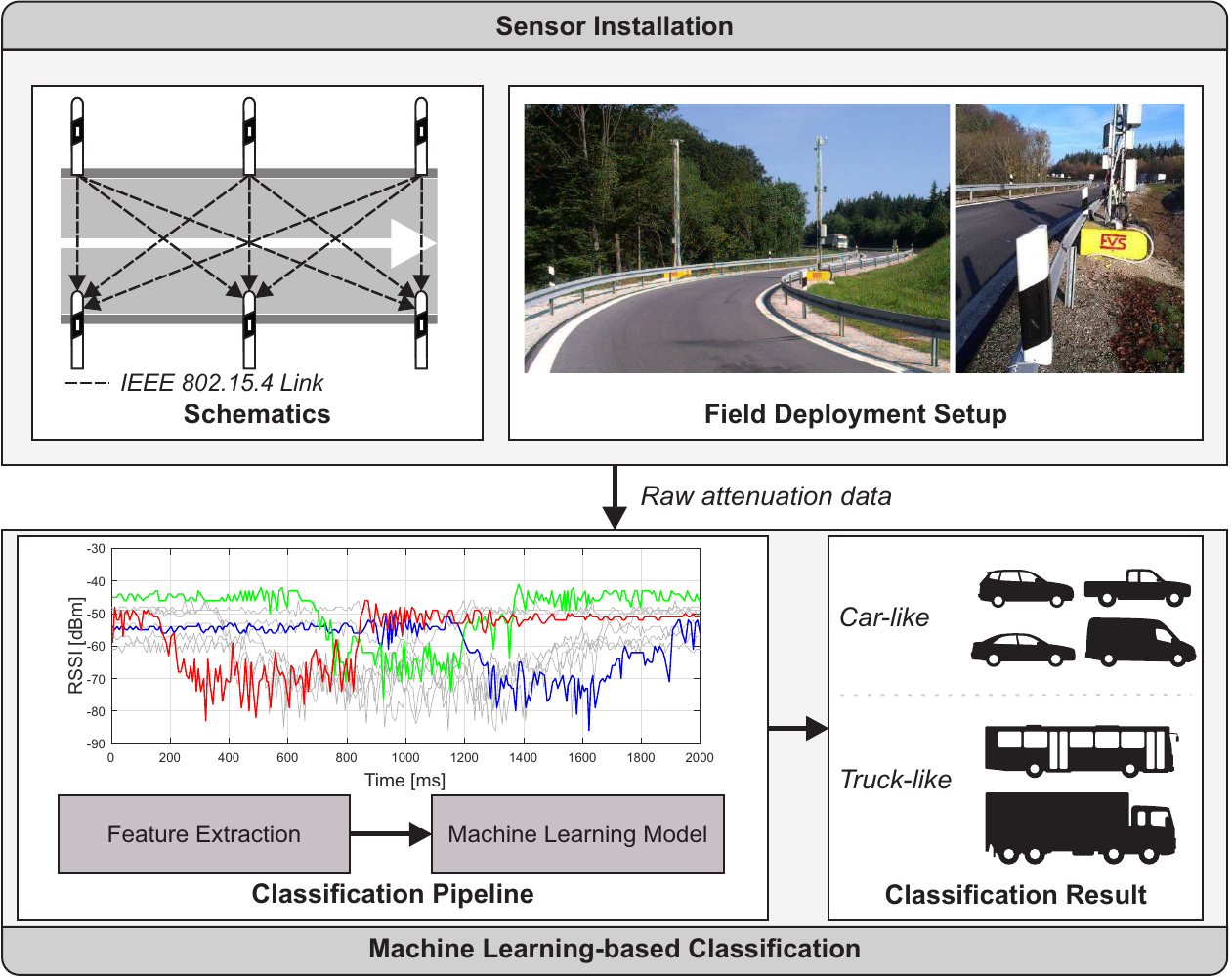}{System model for radio-fingerprinting-based vehicle classification.}{fig:ffws_architecture}
%
%
Fig.~\ref{fig:ffws_architecture} shows the system model for the radio-fingerprinting-based vehicle classification system. For the data acquisition, an experimental deployment of the proposed system at the entrance of a rest area on the German highway A9 within an official test field by the German Federal Ministry of Transport and Digital Infrastructure is used. During the training phase, the measured nine-dimensional time series traces are labeled manually based on camera images. 
%
%
The classification can be performed directly on the raw values as well as on a reduced feature set, which uses analytical properties of the different signals. In \cite{Sliwa/etal/2018e}, a detailed comparison of different machine learning models is provided. 
\basicFig{}{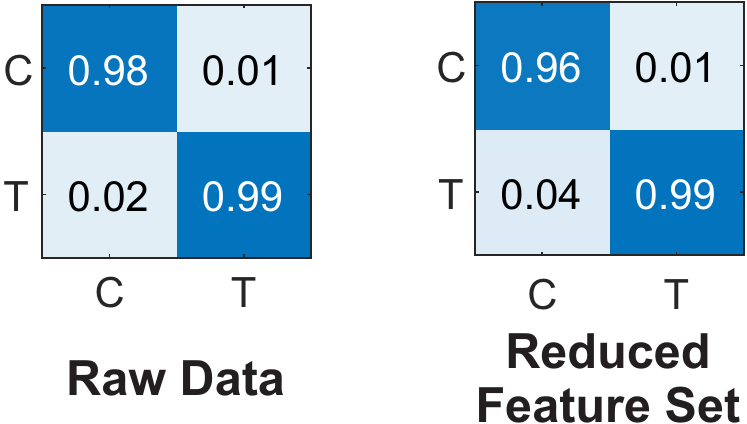}{Confusion matrices for L1/L2 \svm-based binary vehicle classification. (C: Car-like, T: Truck Like)}{fig:ffws_results}{0cm}{0cm}{0.5}
Fig.~\ref{fig:ffws_results} shows the empirical results for L1/L2 \svm-based approach for a data set consisting of more than 2500 real world measurements. For the binary classification task between \emph{car-like} and \emph{truck-like} vehicles, a total classification accuracy of 98.2~\% is achieved.

%
%
With respect to the context of the overall hybrid traffic system, it is assumed that sensor sites exploit their obtained data locally, e.g., for parking space accounting, and contribute themselves to the global traffic knowledge data base through a network connection.
\subsection{Data Transfer: Machine Learning-based Opportunistic Transmission} \label{sec:cat}

In this case study, we present groundwork for resource-efficient transmission of vehicular sensor data for optimizing the coexistence of different resource-consuming cell users. The general approach is to transmit the acquired data in a context-aware manner by exploiting knowledge about the properties of the radio channel. During good channel periods, packet loss is less probable and the resulting data rate is high, therefore the limited resources are only occupied for short time periods. For the opportunistic transmission, acquired sensor data is stored in a local buffer until the expected transmission efficiency is considered sufficient for the whole buffered data.

%
%
\fig{}{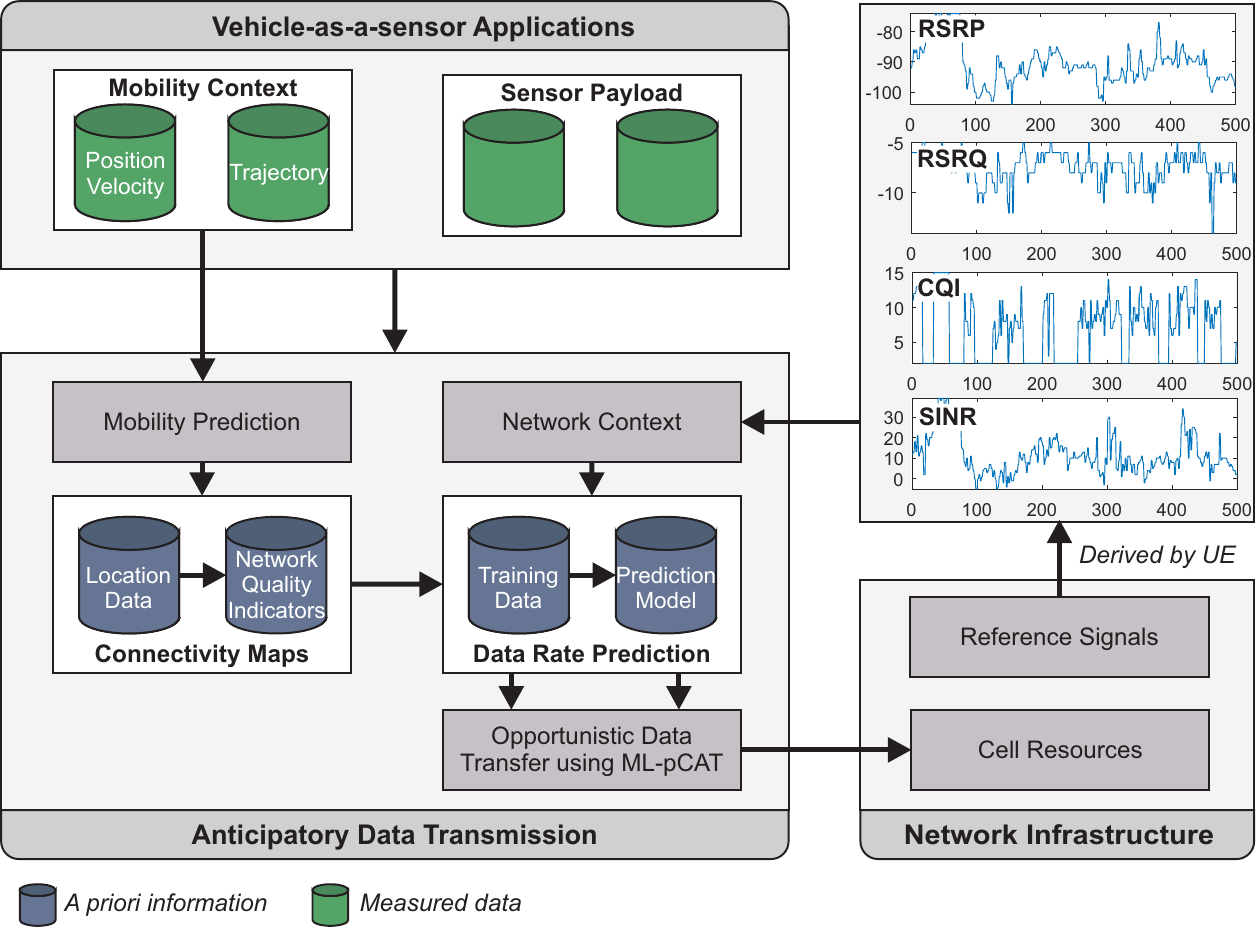}{System model for machine learning-based context-predictive data transmission with \mlpcat.}{fig:cat_architecture}
%
%
\begin{figure*}[b]
	\centering
	\includegraphics[width=.45\textwidth]{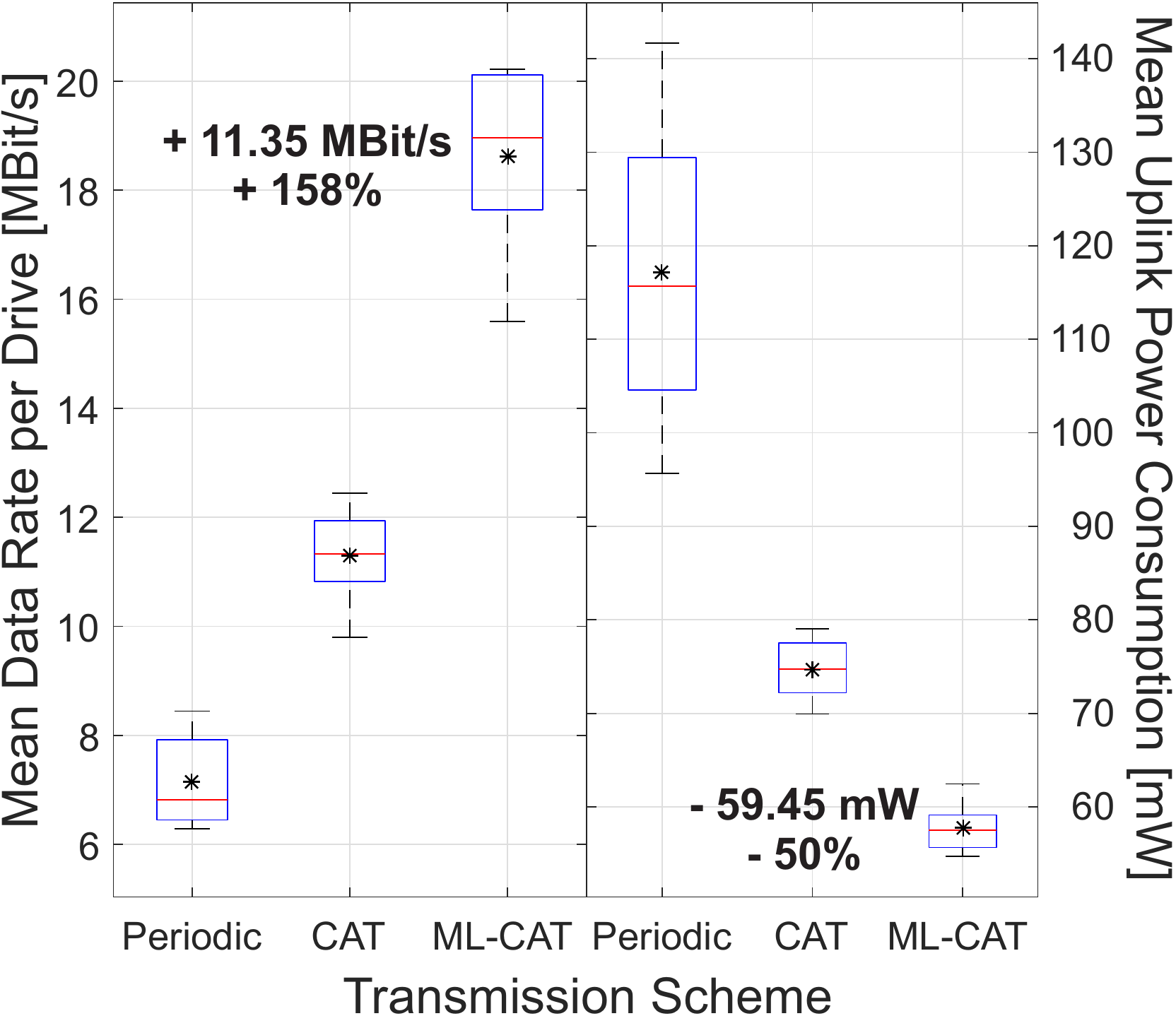}
	\includegraphics[width=.45\textwidth]{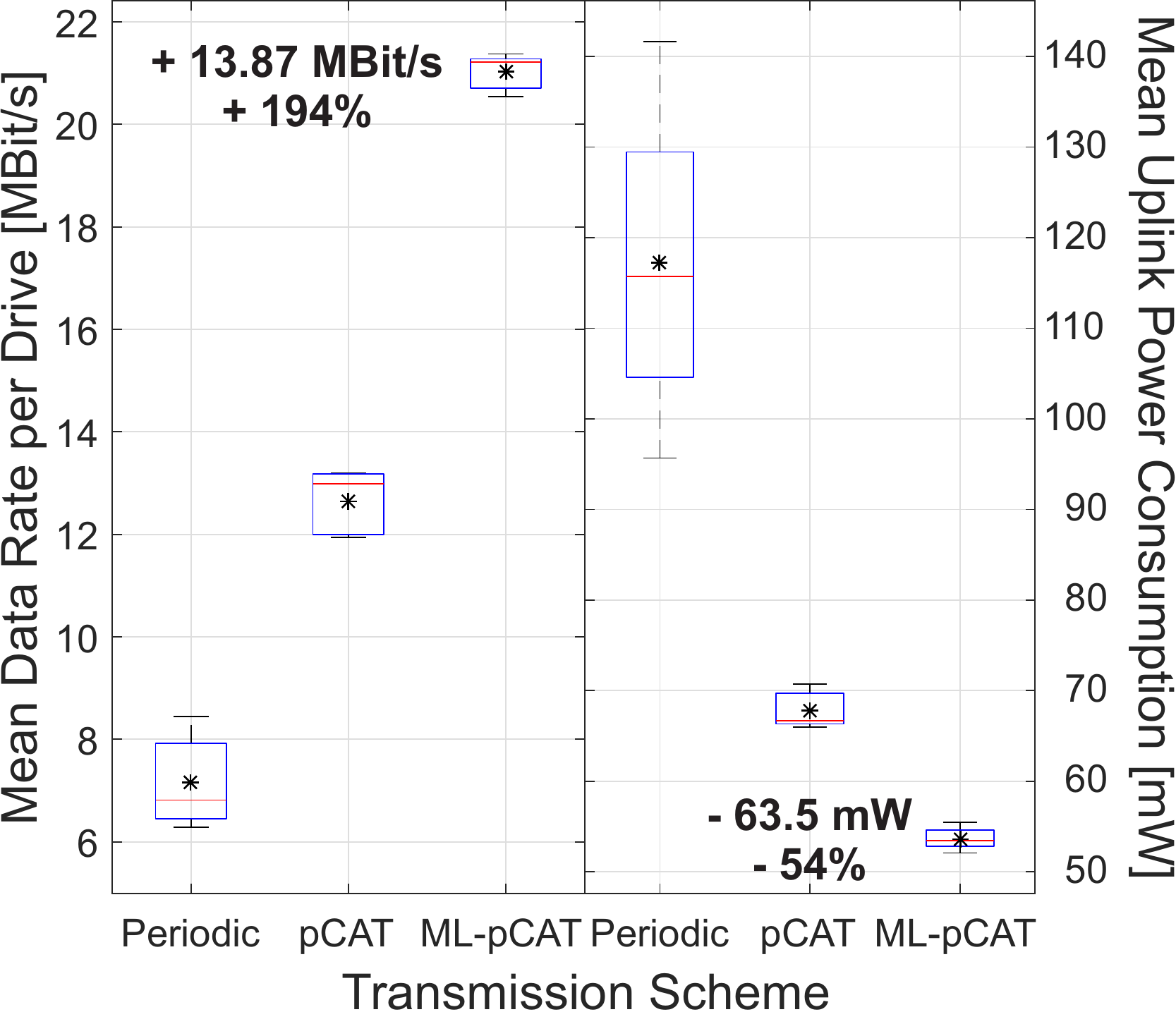}
	\label{fig:ml_cat}
	\caption{Results of real world data transmissions for context-aware (\cat-based) and context-predictive (\pcat-based) approaches. Through integration of machine learning-based data rate prediction both approaches can massively increase the resulting data rate while simultaneously reducing the power consumption of the mobile \ue.}
\end{figure*}

%
%
Based on the transmission schemes \cat and \pcat \cite{Ide/etal/2015a}, which perform data transmissions with respect to the measured \sinr, we developed \mlcat \cite{Sliwa/etal/2018b} and \mlpcat \cite{Sliwa/etal/2018a}, which use machine learning-based data rate prediction as a metric for opportunistic data transfer. The prediction leverages the \lte downlink indicators \rsrp, \rsrq, \sinr, \cqi as well as mobility parameters and payload size to forecast the currently achievable data rate $\Phi(t)$.
For the opportunistic data transmission, \mlcat uses a probabilistic process that computes the transmission probability $p(t)$ based on the currently predicted data rate and its value range defined by $\Phi_{min}$ and $\Phi_{max}$. The weighting exponent $\alpha$ is used to define the preference of high metric values.
%
%
\begin{equation}
	p(t) = \left( \frac{\Phi(t)-\Phi_{min}}{\Phi_{max}-\Phi_{min}} \right)^\alpha
\end{equation}
%
%
%
%
%
With \mlpcat \cite{Sliwa/etal/2018a}, \mlcat is extended to integrate to anticipated future network quality states into the transmission process. Mobility prediction is applied to forecast the future vehicle position and a-priori connectivity map information is exploited to estimate the future network quality based on crowdsensing data.

Fig.~\ref{fig:ml_cat} show the resulting mean data rate and power consumption of \mlcat and \mlpcat in a real world evaluation campsaign using the public cellular network. Data transmissions are performed from a moving vehicle to a cloud-based server, whereas the overall driven distance is more than 2000~km. Further details about the setup and the parametrization are provided in \cite{Sliwa/etal/2018b} and \cite{Sliwa/etal/2018a}.
The proposed \mlcat scheme is compared to naive periodic data transfer (transmission interval 30~s) and the \sinr-based \cat and \pcat schemes.
%
%
It can be seen that the integration of machine learning for network quality assessment leads to massive increases in the end-to-end data rate. The proposed approach is able to implicitly consider complex interdependencies between the different network quality indicators and the transmitted data itself.  
The achieved benefits are even increased by considering the future vehicle context along the vehicle's trajectory with \mlpcat.
%
%
Additionally, the average power consumption of the mobile device is significantly reduced. On the one hand, data transmission are more likely performed at good channel conditions with a low transmission power and a low packet loss probability, on the other hand the transmissions themselves are executed faster due to the higher data rate. Since the applied transmission power has a dominant impact on the overall power consumption \cite{Falkenberg/etal/2018a}, the device battery is used in a more efficient way.

%
%
The results of the considered case study show that anticipatory communication can serve as a powerful mechanism to boost the resource-efficiency of data transmissions from the \ue side without requiring cost-intense extensions of the network infrastructure or the spectrum resources.
\subsection{Data Analysis and Data Exploitation: Modeling and Optimization of Inner City Vehicular Traffic} \label{sec:simu}

%
%
\begin{figure*}[]
	\includegraphics[width=1\textwidth]{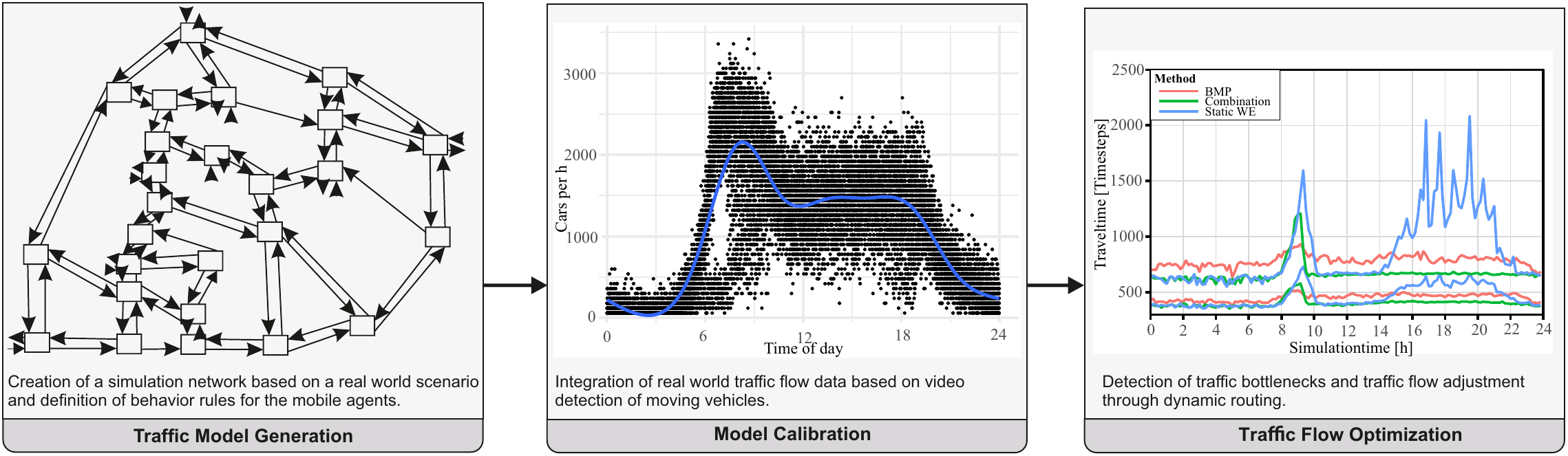}
	\label{fig:ics_architecture}
	\caption{Methodological process for modeling and optimization of realistic traffic scenarios based on real world traffic data. Road network topology information is utilized to generate a simulation setup, which is calibrated using sensor data.}
\end{figure*}
In the final case study, we present a process for modeling and optimization of realistic traffic scenarios based on real world sensor measurements. Although the considered scenario, which is described in further details in \cite{Vranken/etal/2018a}, is focusing on non-automated traffic, it provides valuable methodological groundwork for the proposed overall system model as it involves real world data acquisition, model generation, data analysis for bottleneck detection and data exploitation for traffic flow optimization using dynamic vehicle routing.

%
%
Fig. \ref{fig:ics_architecture} shows the system model for the considered simulation scenario. In the first step, a model for the road network is derived based on real world map knowledge. 
%
%
Real world data of a video-detection system located around and within the German city Düsseldorf is used to calibrate the simulation-parameters of the Break-light-CA-model \cite{Knospe/etal/2000a} and the traffic-inflow.

%
%
For minimizing the global dwell time of the vehicles within the traffic system, the optimization methods \bmp \cite{Kerner/2011a}, \we \cite{Wardrop/52a}, and a newly introduced method, which combines both approaches, are compared. The \bmp aims to minimize the entropy by maximization of the distance to the criticial traffic flow for each detected bottleneck. Although this approach works well for medium to high traffic flows, it causes larger travel times than required for low traffic flow scenarios, resulting in an increased global travel time. The combined approach is able to avoid these situations through integrating the travel times within the model. In the considered scenario, \bmp is able to reduce the average travel time by 23~\% and the combined model further achieves a reduction by 10~\%.

%
%
The results show that dynamic routing methods can successfully utilized for reducing the overall dwell time of vehicles within the traffic system. For the hybrid vehicular traffic, the capabilities for propagating routing information to individual vehicles are limited and depend on the availability of communication technology. Therefore, it needs to be investigated how the established models perform if only a fraction of the vehicles implements the optimization principles.

\section{Conclusion}

%
%
In this paper, we presented a system-of-systems approach for modeling, analysis and optimization of hybrid vehicular traffic system with a focus on data-driven traffic analysis.
%
%
On the long way towards fully-autonomous traffic, the coexistence of systems with heterogeneous automation levels leads to novel challenges for traffic management systems, which need to be addressed by new methods of data acquisition, data transfer, data analysis and data exploitation. Therefore, proposed system-of-systems model relies on close interdisciplinary work of traffic physics, data science and communication engineering.
%
%
Within multiple case studies, we presented groundwork for all main systems aspects and illustrated ways of using the existing road and network infrastructure in more efficient ways.
%
%
In future work, we will continue to transfer the different individual components into the proposed overall system model. For analysis and optimization, simulation models based on measured system properties will be derived. After the simulative optimization phase, adjustments will be fed back into the real world systems.

\section*{Acknowledgment}

\footnotesize
Part of the work on this paper has been supported by Deutsche Forschungsgemeinschaft (DFG) within the Collaborative Research Center SFB 876 ``Providing Information by Resource-Constrained Analysis'', project B4.

\bibliographystyle{IEEEtran}
\bibliography{Bibliography}

\end{document}